\documentclass[twocolumn,twoside,slac_two]{revtex4}
\usepackage{graphicx}
\usepackage{fancyhdr}
\begin{document}

\title{Recent QCD Results}

%

\author{D. Lincoln}
\affiliation{Fermi National, Accelerator Laboratory, P.O. Box 500, Batavia, IL, USA}

\begin{abstract}
The study of the inelastic scattering of hadrons has progressed in the last decade.  With the availability of high-statistics data sets from HERA and the Tevatron, our understanding of high energy and high jet multiplicity events has become rather precise.  In this Proceedings, I present an overview of recent jet-only results, as well as measurements of events which combine both jets and a $W$ or $Z$ boson.  
\end{abstract}

\maketitle

\thispagestyle{fancy}


\section{Introduction}
The study of the interactions between the quark and gluon constituents of hadrons has evolved over the past several decades.  Experiments once had large systematic and statistical uncertainties and theoretical predictions used only leading-order perturbation theory.  However our understanding has considerably improved and precise measurements and calculations are now available.

The theory of the strong interactions, called Quantum ChromoDynamics or QCD \cite{theory_pQCD}, is a very interesting one in its own right.  In addition, because these kinds of interactions dominate at a hadron collider, they form a substantial background for other interesting possible physical processes, including top quark and Higgs boson production, as well as other potential new pheneomena, such as models incorporating the principle of supersymmetry.

In these proceedings, we present recent QCD results, focusing predominantly on data taken at the Fermilab Tevatron.  These subjects include simple jet production, as well as jet production in association with electroweak bosons. 

\section{Accelerator and Detectors}
The Fermilab complex accelerates protons and antiprotons and collides them in their joint center of mass frame at an energy of 1.96 TeV.  These collisions are recorded by the D\O\ \cite{overview_d0} and CDF \cite{overview_cdf} detectors, two large, multi-purpose, detectors located at the Tevatron.  Each detector consists of an inner tracker, composed of a silicon vertex detector and a detector of coarser granularity.  The tracking volume is surrounded by a superconducting solenoid and calorimetry.  The entire detector is enclosed by a magnetized muon tracking detector.  The instantaneous luminosity is as much as $3\times 10^{32}$ cm$^{-2}$ s$^{-1}$ and each experiment has recorded approximately 6 fb$^{-1}$.  The various results presented here make use of 0.3 - 2.5 fb$^{-1}$ of data.

\section{Jet algorithms}
High energy jets are the debris of hadron interactions, which are often modelled as the hard scatter of partons from within the hadrons.  In order to compare experimental measurements (which involve many particles) to theoretical calculations (which generally involve very few), an algorithm is necessary that (a) integrates theoretically-intractable phenomena and (b) is valid over at all levels: parton, particle and detector.

There are two classes of jet-finding algorithms that are used, the cone-based algorithm and some sort of recombination algorithm.  In this proceedings, the results mostly use a cone-based algorithm \cite{theory_cone} which iteratively combined energy with in a cone of radius $R < \sqrt{\Delta \phi^2 + \Delta y^2}$ where $\phi$ is the azimuthal angle, $y = 0.5 \ln [(E+p_z)/(E-p_z)]$ is the rapidity, and $z$ is the direction of the proton beam.  Depending on the analysis, a cone size of $R = 0.7$ or $R = 0.5$ is used.  The seeds for the algorithm were provided by towers in the calorimeter or individual particles or partons.  In order to stabilize the algorithm against low energy emission, the midpoint between reconstructed jets provided an additional set of seeds.

A second algorithm used in these proceedings is the $k_T$ algorithm \cite{theory_kt_1, theory_kt_2}.  This successive recombination algorithm uses all pairs of seeds to calculate $d_{i,j} = \min(p_{T,i}^2,p_{T,j}^2)(\Delta \phi^2 + \Delta y^2)/D^2$, with $D=0.7$ in this proceedings and the $p_{T,i}$ is the transverse momentum of the $i^{\rm th}$ seed.  This is then ordered in value and compared to the list of the transverse momentum squared of the remaining seeds ($d_i = p_{T,i}^2$).  If the minimum is a $d_i$, this it is declared to be a jet and removed from further consideration.  If the minimum is one of the $d_{i,j}$, the two are combined and the process repeated.  This algorithm is intrinsically safe to infrared emission.

\section{Jet-only measurements}
\subsection{Inclusive jet $p_T$ cross section}

\begin{figure}[hf]
\centering
\includegraphics[width=80mm]{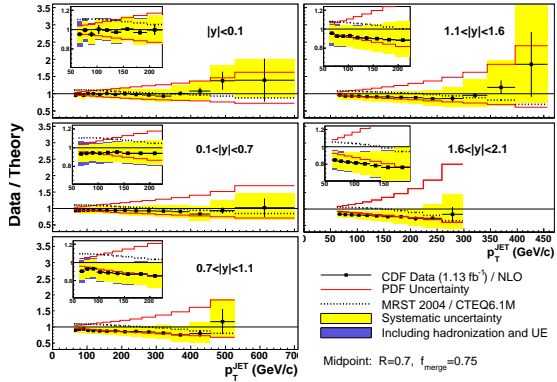}
\caption{CDF: Ratio of measurement to data in the cone inclusive spectrum.} \label{figure_cdf_inclusive_cone}
\end{figure}

\begin{figure}[hf]
\centering
\includegraphics[width=80mm]{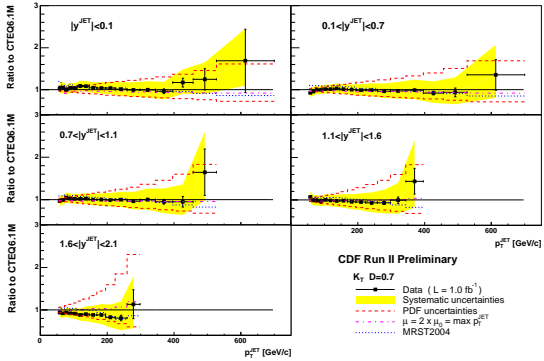}
\caption{CDF: Ratio of measurement to data in the $k_T$ inclusive spectrum.} \label{figure_cdf_inclusive_kt}
\end{figure}

CDF has published the inclusive jet cross section for both the cone \cite{cdf_incl_cone} and $k_T$ \cite{cdf_incl_kt} algorithms.  Both analyses present the data in five bins in rapidity, covering a range to $|y| < 2.1$.  The results include approximately 1 fb$^{-1}$ of data.  The measurement was corrected to the particle level, as were the theoretical calculations.  The systematic uncertainties are substantial and are dominated by the jet energy scale correction.  Figures \ref{figure_cdf_inclusive_cone} shows the ratio of the cone measurement to theory, while figure \ref{figure_cdf_inclusive_kt} shows the corresponding measurement using the $k_T$ algorithm.

\begin{figure}[hf]
\centering
\includegraphics[width=80mm]{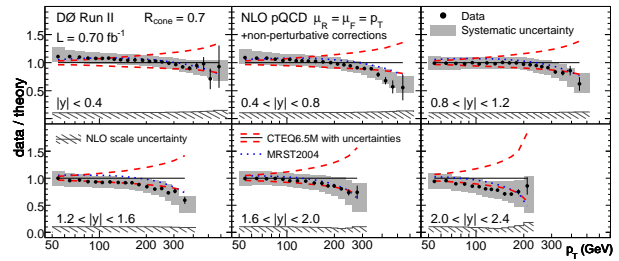}
\caption{D{\O}: Ratio of measurement to data in the cone inclusive spectrum.} \label{figure_d0_inclusive_cone}
\end{figure}

The D{\O} experiment has published \cite{d0_incl_cone} a measurement of the inclusive jet cross section as a function of jet $p_T$ in six rapidity bins, covering a range of $|y| < 2.4$.  Exploiting the liquid argon/uranium calorimetry, along with a detailed understanding of the response of the calorimeter to both quarks and gluons, they were able to measure these quantities with unprecedented precision; approximately 30-50\% smaller than comparable CDF measurements.  Figure \ref{figure_d0_inclusive_cone} shows the ratio of the measurement to next-to-leading order theoretical calculations, using the CTEQ 6.5M structure functions.  The data clearly is below the calculations at high jet $p_T$.  This observation is taken as evidence that the PDFs used in this analysis might have too much momentum carried by gluons with a large fraction of the beam momenta.  This data was included in a recent PDF extraction \cite{MSTW2008}.  

\subsection{Dijet mass cross section}

Both D\O\ and CDF have published analyses studying the dijet mass spectrum, based on the same data set as was used for the inclusive jet analysis.  Both experiments limited themselves only to measurements using the cone algorithm.  The CDF measurement \cite{cdf_dijet_mass} was restricted to a single rapidity range extending to $|y| < 1.0$, while D{\O}'s measurement \cite{d0_dijet_mass} consisted of six bins in rapidity, extending to $|y| < 2.4$.

\begin{figure}[hf]
\centering
\includegraphics[width=80mm]{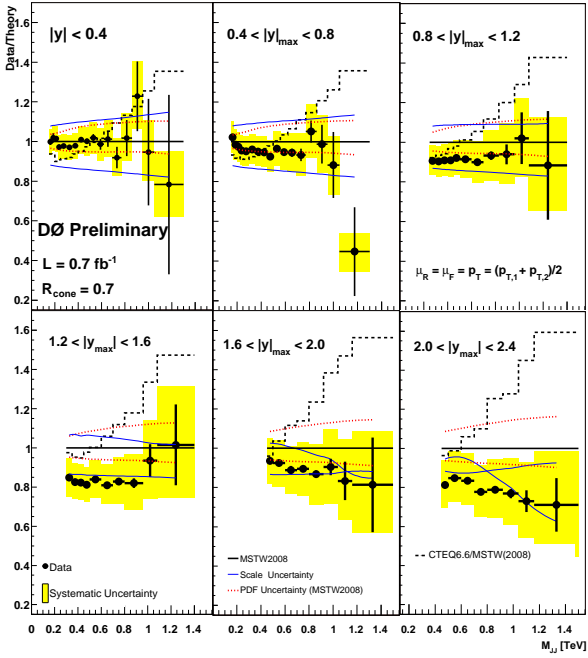}
\caption{D{\O}: Ratio of measurement to data in the inclusive dijet mass spectrum.} \label{figure_d0_dijet_mass}
\end{figure}

\begin{figure}[hf]
\centering
\includegraphics[width=80mm]{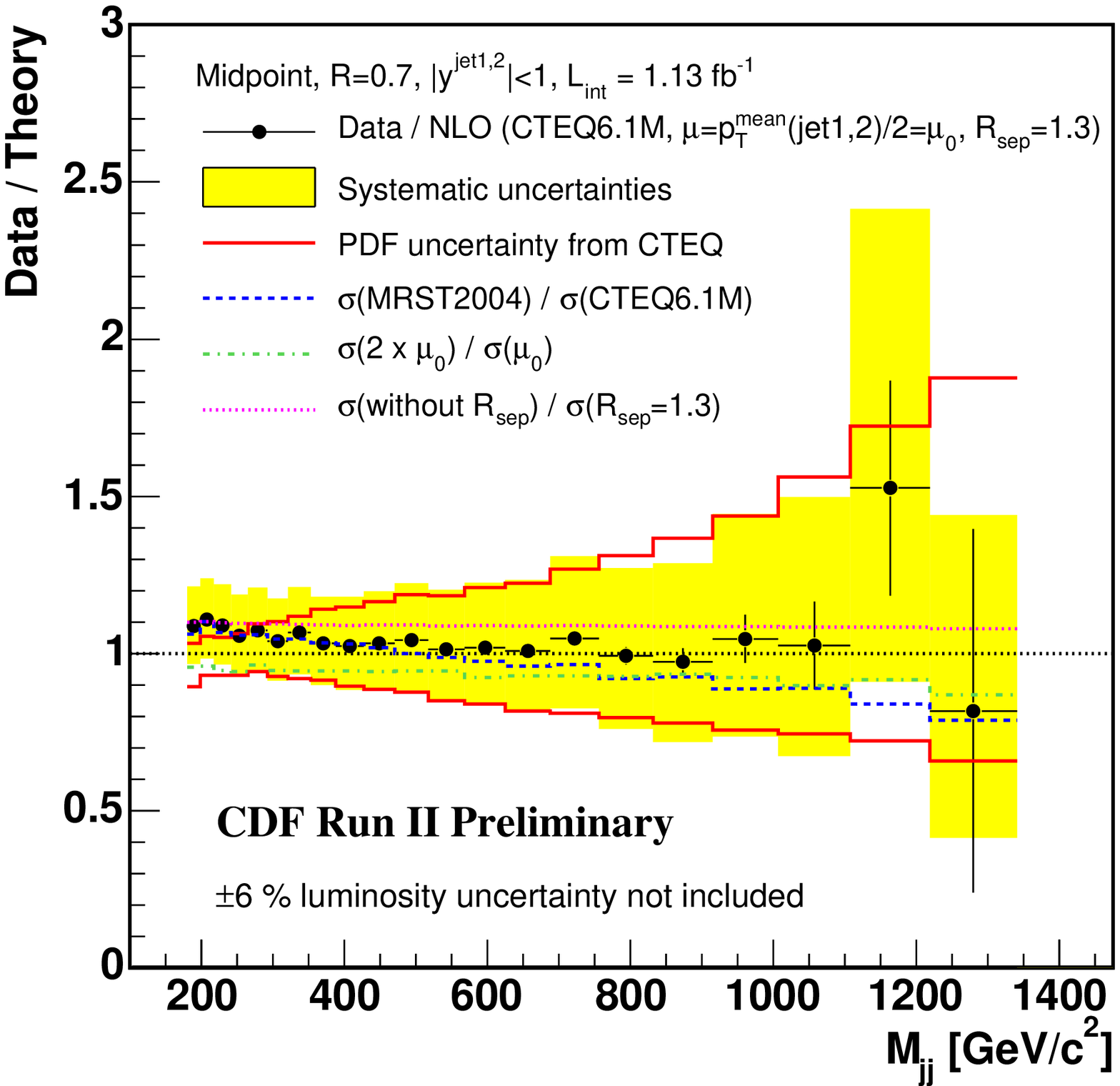}
\caption{CDF: Ratio of measurement to data in the inclusive dijet mass spectrum.} \label{figure_cdf_dijet_mass}
\end{figure}

Figures \ref{figure_d0_dijet_mass} and \ref{figure_cdf_dijet_mass} tell a story comparable to the jet inclusive $p_T$ measurements.  However, for this analysis,  the D{\O} measurement compares to theory utilizing the MSTW2008 PDFs.  These PDFs include Tevatron Run II data (including the D{\O} inclusive jet cross section which is based on the same data set), but no Tevatron Run I data.  

\begin{figure}[hf]
\centering
\includegraphics[width=80mm]{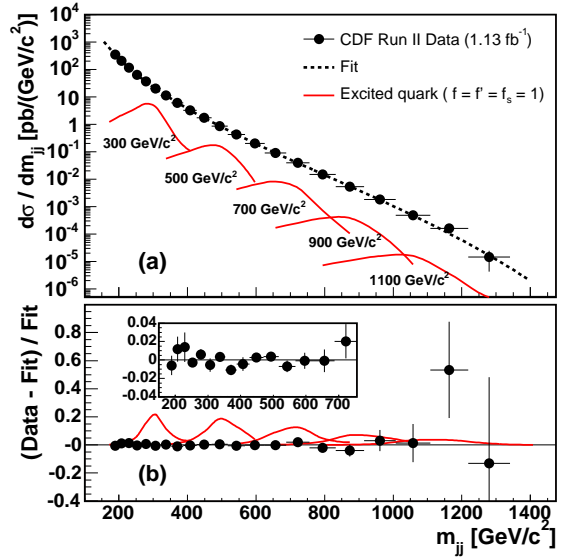}
\caption{CDF: Comparison of data to a fit with dijet resonance signals overlaid.} \label{figure_cdf_dijet_resonance}
\end{figure}

In addition, CDF has used its dijet mass distribution (shown in Figure \ref{figure_cdf_dijet_resonance}) to search for dijet resonances.  They have set limits on axigluon/coloron, E6 diquark, color octet (techni-$\rho$) and excited quarks.  Table \ref{table_cdf_resonance_limits} shows the published limits and a detailed description of the various models can be found in the references.

\begin{table}
\begin{tabular}{cc}
Model & Excluded Mass (GeV) \\
axigluon, coloron & 260-1250 \\
E6 diquark & 260-630 \\
color octet (Techni-$\rho$) & 260-110 \\
excited quark & 260-870 
\end{tabular}
\caption{CDF: Limits set on new physics using their dijet mass resonance search.}
\label{table_cdf_resonance_limits}
\end{table}

\subsection{Dijet angular distribution}

Both D\O\ \cite{d0_dijet_chi} and CDF \cite{cdf_dijet_chi} have published analyses studying the dijet angular distribution as a function of the dijet mass.  This particular analysis is attractive in that it is sensitive to searches for new physics and is relatively insensitive to uncertainties in the PDFs.

The essence of this approach is that standard QCD scattering occurs at small angles with respect to the beam direction, while new phenomena is expected to occur at larger angles.  The variable used in this analysis is $\chi = \exp(|y_1 - y_2|)$ where $y_1$ and $y_2$ are the rapidities of the two leading jets, ordered in jet $p_T$.

\begin{figure}[hf]
\centering
\includegraphics[width=80mm]{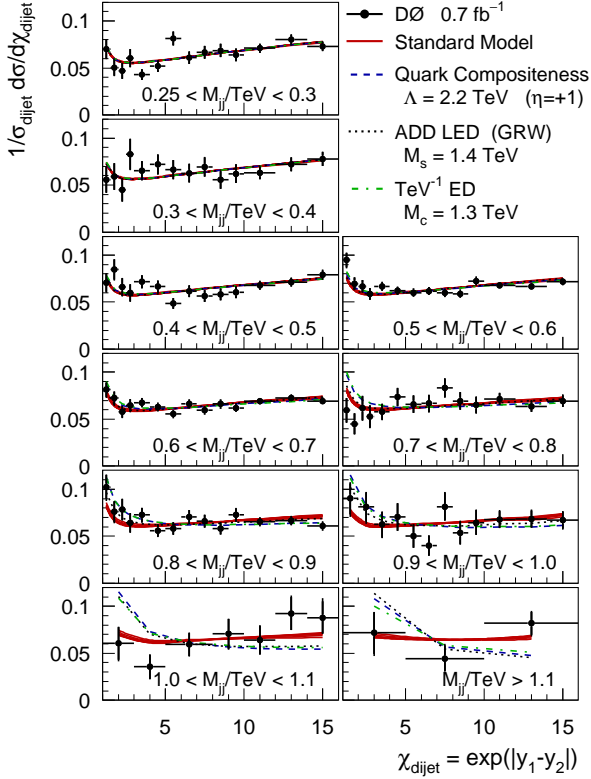}
\caption{D{\O}: Dijet angular distribution for ten bins of dijet mass.} \label{figure_d0_angular}
\end{figure}

Figure \ref{figure_d0_angular} shows D{\O}'s published result of the normalized $\chi$ distribution for eight bins, including the first data in this variable for a dijet mass exceeding 1.1 TeV.  The data is compared to several models, including quark compositeness, and two models of extra dimensions.  The data agrees well with the standard model and stringent limits are set on these speculative processes.  For quark compositeness, limits can be set on the compositeness scale, which must exceed 2.91(2.97) TeV for constructive(destructive) interference.

\begin{figure}[hf]
\centering
\includegraphics[width=80mm]{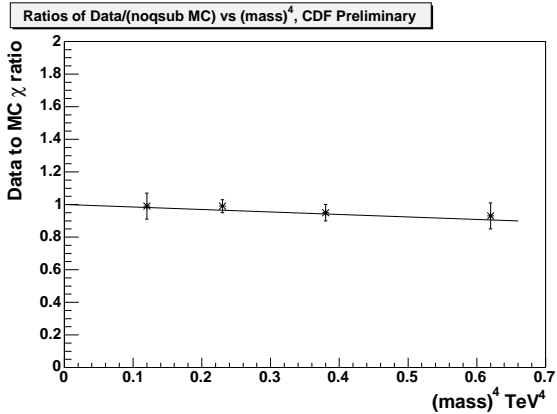}
\caption{CDF: Double ratio of the integral of the $\chi$ distribution for $(1 \leq \chi \leq 10)/(15 \leq \chi \leq 25)$ for data over monte carlo predictions.} \label{figure_cdf_angular}
\end{figure}

Figure \ref{figure_cdf_angular} shows CDF's ratio of data to Monte Carlo predictions for the integral of the $\chi$ distribution for $(1 \leq \chi \leq 10)/(15 \leq \chi \leq 25)$.  The data is in good agreement with QCD prediction, with no hint of physics outside the current model.

\section{Jets + Vector Bosons}
While the production of jets has historically been the main focus of QCD analyses, events which contain both jets and vector bosons are an ideal sample to investigate physical processes with relatively low cross-sections.  The existence of the vector boson greatly suppresses the backgrounds.  Typical physical processes of interest include top quark production, Higgs boson production and various new phenomenon models.

Both D\O\ and CDF have made extensive measurements of events containing vector bosons and up to four jets. 

\subsection{$W$ + jets}

CDF has reported \cite{cdf_w_jets} a study of $W + \geq n$ jet production using 320 pb$^{-1}$ of data where $n$ is up to 4. This study used the electron decay mode of the $W$ boson.  The jets found in this analysis used a cone-based algorithm with $R = 0.4$.  It required the electron to have $p_T > 20$ GeV and $|\eta_e| < 1.1$.  The neutrino was required to have $p_T > 30$ GeV and the transverse mass of the $W$ was $m_T(W) > 20$ GeV.  The data was compared to both NLO and LO calculations, with the details found in the reference.

As approximately expected, data was found to be in good agreement with NLO and to have good shape agreement with LO.  Leading order calculations appear to under-predict the data by about 40\%.

\begin{figure}[hf]
\centering
\includegraphics[width=80mm]{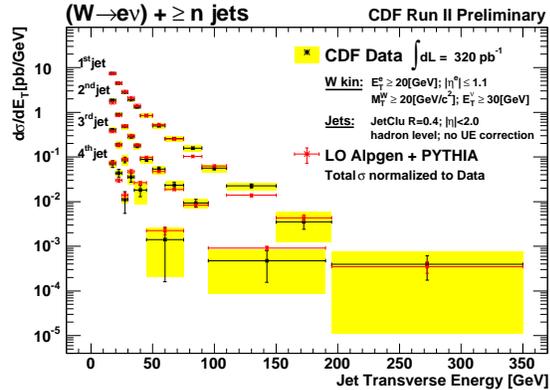}
\caption{CDF: Jet $p_T$ distribution in $W$ + jet events for up to three jets.} \label{figure_cdf_w_jets}
\end{figure}

\subsection{$Z$ + jets}
\begin{figure}[hf]
\centering
\includegraphics[width=80mm]{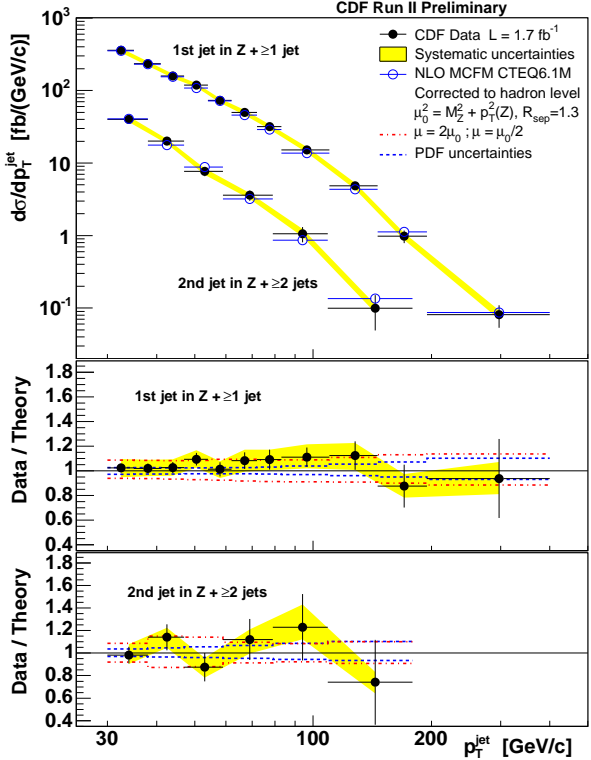}
\caption{CDF: Jet $p_T$ distribution for $Z$ + jet events, for the two leading jets.  In addition, a ratio is formed with NLO theory.} \label{figure_cdf_z_jets}
\end{figure}

CDF also has a published an analysis \cite{cdf_z_jets} describing the behavior of events which have a $Z$ boson and 1-3 jets.  The paper compares the data to NLO theory and good agreement is found both in the jet multiplicity distribution and in the jet $p_T$ distribution (shown in figure \ref{figure_cdf_z_jets}).  The systematic uncertainties are on the order of 10\%.  Leading order calculations have the right shape, but the wrong normalization.

D\O\ has published analyses of $Z$ + $n$ jet events, with both electron \cite{d0_z_jets_1} and muon \cite{d0_z_jets_2} decay channels.  The data is compared to both LO and NLO calculations \cite{MCFM} and leading order (PYTHIA \cite{pythia}, HERWIG \cite{herwig}), and higher order tree-level (ALPGEN \cite{alpgen}, SHERPA \cite{sherpa}) generators, both coupled with parton showers.  In the first paper \cite{d0_z_jets_1}, the jet $p_T$ is presented in events with up to 3 jets.  Both HERWIG and PYTHIA show significant shape and normalization differences when compared to the data.  In contrast, both ALPGEN and SHERPA, which combine tree-level calculations with parton shower enhancements are in reasonable agreement with the shape of the data distributions.  The residual normalization differences suggest residual large scale uncertainties.  However, for a specific choice of scale, the normalization of either generator can be made to agree with the data.

\begin{figure}[hf]
\centering
\includegraphics[width=80mm]{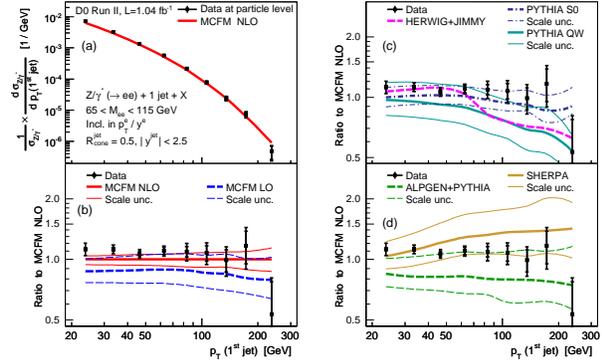}
\caption{D{\O}: Representative plot of the leading jet $p_T$ distribution for $Z$ + jet events. The cited article \cite{d0_z_jets_1} has similar distributions for up to 3 jets.} \label{figure_d0_z_jets_1}
\end{figure}

\begin{figure}[hf]
\centering
\includegraphics[width=80mm]{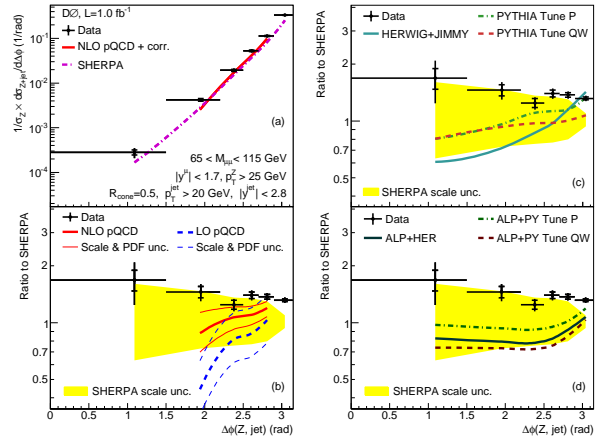}
\caption{D{\O}: Representative plot of the $\Delta \phi$ distribution between the $Z$ boson and the leading jet. The cited article \cite{d0_z_jets_2} has similar distributions for other angular variables.} \label{figure_d0_z_jets_2}
\end{figure}

A second D\O\ $Z$ + jets analysis \cite{d0_z_jets_2} uses the muon decay channel and investigates the angular correlations of the $Z$ boson and the leading jet.  In this analysis, the distributions $\Delta \phi(Z,{\rm jet})$, $\Delta y(Z,{\rm jet})$, and $y_{\rm boost}(Z,{\rm jet})$ are presented.  Figure \ref{figure_d0_z_jets_2} shows the $\Delta \phi(Z,{\rm jet})$ result.  The summary conclusions of the paper is that ALPGEN and SHERPA are superior generators, although there remain scale dependences in the theory that it is hoped that tuning to this data will reduce.

\section{The LHC Era}
With the turn-on of the LHC, it is naturally expected that new limits will be set for the higher jet $p_T$ and dijet mass scales.  However, in order to be competitive they will need to achieve systematic uncertainties.  In order to achieve the uncertainties that D\O\ quotes here required several years of study of their calorimetry.  It is expected that an extended period of time will be needed by both CMS and ATLAS.

Early QCD results from the LHC will be the charged particle spectra within jets and jet measurements that are relatively insensitive to jet energy corrections, for instance the $\Delta \phi$ distribution between the two leading jets. 

\section{Summary}
The study of QCD has now come of age.  Very precise measurements are available, as are calculations that can be as high as NNLO in precision.  The magnitude of the high-$x$ gluon contribution to the PDFs remains a question of some concern and it is hoped that the precise measurements of the Tevatron will reduce the uncertainty associated with this phenomenon.  The study of $W/Z$ + jets events is crucial to reduce the background systematic uncertainties for Higgs boson searches, as well as other new phenomenon.

The era of the LHC is upon us.  There is considerable excitement among the field as to what lessons the LHC data will provide.  Within two years of turn on, it is quite possible that the extended energy range will have revealed new information in our study of the fundamental building blocks of the universe.

\bigskip 

\end{document}